\documentclass[
    ,final            
  ]
  {aipproc}

\layoutstyle{6x9}
\usepackage{amssymb}

\begin{document}

\title{Observing Neutral Hydrogen Above Redshift 6: The ``Global'' Perspective}


\author{Judd D. Bowman}{
  address={California Institute of Technology, Pasadena, CA 91125, USA}
  ,altaddress={Hubble Fellow}
}

\author{Alan E. E. Rogers}{
  address={Haystack Observatory, Massachusetts Institute of Technology, Westford, MA 01886, USA}
}

\author{Jacqueline N. Hewitt}{
  address={Kavli Institute for Astrophysics and Space Research, Massachusetts Institute of Technology, Cambridge, MA 02139, USA}
}

\begin{abstract}
Above redshift 6, the dominant source of neutral hydrogen in the Universe shifts from localized clumps in and around galaxies and filaments to a pervasive, diffuse component of the intergalactic medium (IGM).  This transition tracks the global neutral fraction of hydrogen in the IGM and can be studied, in principle, through the redshifted 21 cm hyperfine transition line.  During the last half of the reionization epoch, the mean (global) brightness temperature of the redshifted 21 cm emission is proportional to the neutral fraction, but at earlier times ($10 < z < 25$), the mean brightness temperature should probe the spin temperature of neutral hydrogen in the IGM.  Measuring the (of order 10~mK) mean brightness temperature of the redshifted 21 cm line as a function of frequency (and hence redshift) would chart the early evolution of galaxies through the heating and ionizing of the IGM by their stellar populations. Experiments are already underway to accomplish this task or, at least, provide basic constraints on the evolution of the mean brightness temperature.  We provide a brief overview of one of these projects, the Experiment to the Detect the Global EOR Signature (EDGES), and discuss prospects for future results.
\end{abstract}

\maketitle


\section{Introduction}

Understanding the complete history of hydrogen gas in the Universe from recombination to today is one of the most fundamental topics in astrophysics and is central to current studies in cosmology and galaxy evolution.  The earliest probe of baryonic matter comes from the cosmic microwave background (CMB), which illuminates the conditions when hydrogen gas first formed during recombination at $z\approx1000$.  But despite this initial data point, our knowledge is virtually empty until the ``modern'' era of quasars, galaxies, and large-scale structure below $z\lesssim6$, when hydrogen gas is primarily localized in galaxies.  During the intervening period ($1000 \gtrsim z \gtrsim 6$), most hydrogen gas resides in the intergalactic medium (IGM), where it is difficult to observe with existing techniques.  During this period, the spin temperature and ionization states of the IGM encode the process of baryon collapse and the properties of the earliest stars, galaxies, and quasars (see \cite{2006PhR...433..181F} for a recent review).  New instruments are under development to extract this information by detecting emission or absorption from the redshifted 21 cm hyperfine transition line of diffuse neutral hydrogen in the high-redshift IGM relative to the CMB.   At the target redshifts, the 21 cm line has been redshifted from its characteristic (rest-frame) frequency of 1420~MHz to 200~MHz or less.  At these low radio frequencies, Galactic synchrotron radiation dominates the sky and is a factor of $10^4$ to $10^9$ more intense than the predicted $\sim10$~mK redshifted 21 cm signal.

\section{Experimental Approach and Preliminary Results}

One approach to extracting information about hydrogen gas in the high-redshift IGM is to measure the mean redshifted 21 cm contribution to the all-sky radio spectrum as a function of frequency (and hence redshift).  Such a measurement should yield the redshift evolution of the global spin temperature and ionization fraction of the hydrogen gas in the IGM above $z>6$ and may provide the first constraints on the history of hydrogen during this period.  In this paper, we briefly describe the preliminary results and future prospects for one current effort, the Experiment to Detect the Global EOR Signature (EDGES).  A detailed description of this project can be found in \cite{edges_eor}.

Determining the $< 30$~mK redshifted 21~cm contribution to the radio spectrum requires separating the signal from the foreground spectrum at better than one part in $10^4$. In principle, this may be accomplished by taking advantage of the differences between the spectra of the Galactic and extragalactic foregrounds and the anticipated redshifted 21 cm contribution.  The astrophysical foregrounds are all characterized by smooth power-law spectra, whereas the redshifted 21 cm signal is expected to have up to three rapid transitions \citep{1999A&A...345..380S, 2004ApJ...608..611G, 2006MNRAS.371..867F} corresponding to the cooling of the IGM, followed by the heating of the IGM, and finally reionization.  This difference in spectral structure can be exploited with an experiment that constrains the smoothness of the spectrum to very high precision. The primary need is to reduce any instrumental or systematic contributions to the measured power spectrum that vary rapidly with frequency, since these could be confused with a sharp feature in the spectrum due to a fast reionization of the IGM. Such contributions could be due to terrestrial transmitters, reflections of receiver noise or sky noise from nearby objects, undesirable resonances within the electronics or enclosures, or spurious signals introduced by the digital sampling system.

In order to address these challenges, EDGES consists of three simple modules: 1) an antenna, 2) an amplifier and comparison switching module, and 3) an analog-to-digital conversion and storage unit. The design of the system features several novel elements. The antenna is a ``fat'' dipole-based design derived from the four-point antenna of \cite{fourpoint2}. It is compact and planar in order to reduce self-reflections and placed over a conducting mesh that rests directly on the ground in order to eliminate reflections from the ground and to reduce gain toward the horizon. The amplifier module is connected directly to the antenna without transmission cables. This reduces the impact of reflections within the electrical path of the instrument. The amplifier chain is connected through a voltage controlled switch in order to insert a comparison source to subtract glitches and spurious instrumental signals in the measured sky spectrum. Analog-to-digital conversion is accomplished with an off-the-self Acqiris AC240 8-bit digitizer contained on a CompactPCI card connected to a host computer.  The broadband spectrometer employs the FPGA code of \citet{2005A&A...442..767B} and a Blackman-Harris window function is used to improve the isolation between neighboring frequency channels.

The EDGES system was deployed at the radio-quiet Mileura Station in Western Australia from 29 November through 8 December, 2006.  These dates were chosen such that the Galactic center would be below the horizon during most of the night, keeping the system temperature as low as possible for the measurements.  The system was operated on 8 consecutive nights during the deployment.  Over 30 hours of relevant drift scans were obtained.  To look for small deviations from the smooth foreground spectrum, a seventh-order polynomial was fit to the measured spectrum between 130 and 190~MHz and subtracted.

The rms level of the systematic contributions to the measured spectrum was found to be $T_{rms} = 75$~mK, a factor of $\sim3$ larger than the maximum expected redshifted 21 cm feature that would result from a rapid reionization.  The residuals were due to instrumental contributions and not thermal noise. The largest variations were the result of the 166~MHz PCI-bus clock of the AC240 and computer.  By simulating the sky spectrum, we can determine the rms of the residuals that should remain following the polynomial fit for different reionization scenarios. Comparing the rms of the residuals in the models to the 75~mK rms of the initial measurements gives a good estimate of the region of parameter space ruled out so far. The best constraint, in the case of a nearly instantaneous reionization at $z = 8$, is that the redshifted 21 cm contribution to the spectrum is not greater than about 450~mK before the transition.

The initial measurements with the EDGES system provide a solid foundation for assessing the potential of such a system to extract meaningful information about hydrogen gas in the high-redshift IGM.  With a factor of 20 improvement in the systematic residuals, EDGES should be able to constrain the duration of reionization to $\Delta z > 2$ or better.  Further improvements in the system and longer observing runs could begin to probe the transition from absorption to emission predicted for the redshifted 21 cm when the first luminous sources begin to heat the IGM, as well as limit contributions to the heating of the very high-redshift IGM from exotic physics such as the evaporation of primordial black holes.



%


\begin{theacknowledgments}
  This work was supported by the Massachusetts Institute of Technology, School of Science, and by the NSF through grant AST-0457585.  JDB is supported by NASA through Hubble Fellowship grant HF-01205.01-A awarded by the Space Telescope Science Institute, which is operated by the Association of Universities for Research in Astronomy, Inc., for NASA, under contract NAS 5-26555.
\end{theacknowledgments}



\end{document}